# Spin Electronics

# Capacitively Driven Global Interconnect with Magnetoelectric Switching Based Receiver for Higher Energy Efficiency

Zubair Al Azim[1], Akhilesh Jaiswal[1], Indranil Chakraborty[1], and Kaushik Roy[1]*

[1] *Department of Electrical and Computer Engineering, Purdue University, West Lafayette, IN 47907*
* *Fellow, IEEE*

*Abstract*—We propose capacitively driven low-swing global interconnect circuit using a receiver that utilizes magnetoelectric (ME) effect induced magnetization switching to reduce the energy consumption. Capacitively driven wire has recently been shown to be effective in improving the performance of global interconnects. Such techniques can reduce the signal swing in the interconnect by using a capacitive divider network and does not require an additional voltage supply. However, the large reduction in signal swing makes it necessary to use differential signaling and amplification for successful regeneration at the receiver, which add area and static power. ME effect induced magnetization reversal has recently been proposed which shows the possibility of using a low voltage to switch a nanomagnet adjacent to a multi-ferroic oxide. Here, we propose an ME effect based receiver that uses the low voltage at the receiving end of the global wire to switch a nanomagnet. The nanomagnet is also used as the free layer of a magnetic tunnel junction (MTJ), the resistance of which is tuned through the ME effect. This change in MTJ resistance is converted to full swing binary signals by using simple digital components. This process allows capacitive low swing interconnection without differential signaling or amplification, which leads to significant energy efficiency. Our simulation results indicate that for $5-10\ mm$ long global wires in IBM $45\ nm$ technology, capacitive ME design consumes $3\times$ lower energy compared to full-swing CMOS design and $2\times$ lower energy compared to differential amplifier based low-swing capacitive CMOS design.

## I. INTRODUCTION

With continuous scaling of CMOS technology, the increasing latency and energy dissipation in a global wire remains a challenging problem [Rabaey 2009, Postman 2012]. Implementing an efficient low-swing interconnect topology could be the key to properly address this issue [Zhang 2000, Krishna 2010]. Capacitively driven global wire has recently been proposed to be a promising technique to reduce the signal swing and subsequently the dynamic energy dissipation in a global interconnect [Ho 2008, Mensink 2010, Walter 2012]. A series coupling capacitor driving a long wire increases the signal bandwidth and reduces the signal swing along the line without the necessity of an additional lower power supply, which is desirable for digital IC design [Ho 2008]. The series capacitance acts as a high-frequency short which increases the wire bandwidth and decreases the propagation delay without the necessity of any repeaters. The series capacitance also creates a capacitive divider network, which reduces the wire signal swing and leads to lower capacitive power loss along the line. However, the largely reduced signal swing makes the signal restoration challenging at the receiving end. Almost all the existing capacitive low-swing designs require the use of differential signaling and amplification for successful signal restoration at the receiver [Ho 2008, Mensink 2010, Walter 2012]. The use of differential signaling doubles the wire area while the use of differential amplifiers leads to static energy dissipation [Postman 2012].

Voltage-induced magnetoelectric (ME) effect has recently been shown to reverse the magnetization of a nanomagnet by applying low voltages to an adjacent multi-ferroic oxide [Ramesh 2014, Heron 2014]. The use of ME effect provides a capacitive write port so that a low voltage signal can directly be used to modify the magnetization [Jaiswal 2017a]. Moreover, ME effect induced magnetization reversal offers improvement in terms of speed and energy in comparison to classic current induced switching [Nikonov 2014, Manipatruni 2015]. The switchable nanomagnet can be used as the free-layer of a magnetic tunnel junction (MTJ), the resistance of which will be tunable thorough the ME effect induced magnetization reversal. Consequently, the ME effect induced switching can provide an efficient mechanism to utilize a low voltage at the interconnect receiving end to tune an MTJ resistance. The MTJ resistance change can be easily converted to full swing binary voltages with the use of simple digital elements. As a result, capacitively driven global wire with an ME effect induced receiver can offer higher energy efficiency since it enables operation without the necessity of differential signaling or amplification.

The rest of this letter is organized as follows. In Section II, we first present the details of ME switching based MTJ device (ME-MTJ [Sharma 2015, Jaiswal 2017b]) including the modeling and simulation framework. In Section III, we present the capacitively driven interconnect circuit using an ME-MTJ receiver. We present our simulation results in Sec IV and compare with existing techniques. Section V concludes the letter.

Corresponding author: Zubair Al Azim (zazim@purdue.edu).

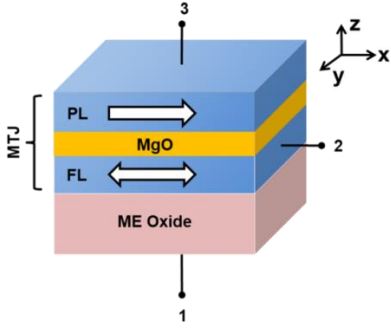

Fig. 1. Schematic of ME-MTJ device. Free Layer (FL) can be reversed by applying a large enough voltage across terminals 1 and 2. State of the FL can be read by using terminals 2 and 3.

Table 1. Device parameters used for our simulations [Jaiswal 2017b]

| Parameters | Value used |
|---|---|
| Magnet length | $45 \times 2.5\ nm$ |
| Magnet width | $45\ nm$ |
| Magnet thickness | $2.5\ nm$ |
| ME oxide thickness | $5\ nm$ |
| MgO thickness | $1.4\ nm$ |
| Saturation magnetization | $1.257 \times 10^3\ A/m$ |
| Gilbert damping constant | 0.03 |
| Interface anisotropy | $1\ mJ/m^2$ |
| ME coefficient | $0.5/3 \times 10^8\ sm^{-1}$ |

## II. ME-MTJ DEVICE AND MODELING

### A. ME-MTJ Device

Magnetoelectric (ME) effect is the modification of magnetization through the application of an electric field [Fiebig 2005, Spaldin 2005]. A nanomagnet in contact with either a single phase or composite multi-ferroic material (such as BiFeO3, BaTiO3 etc.) is the core element in an ME device [Wu 2013, Wu 2015]. The particular ME device under consideration is shown in Fig. 1. It consists of a nanomagnetic free layer (FL) in contact with a multi-ferroic material (ME Oxide) as shown in Fig. 1. A metal contact to the ME oxide, the ME oxide itself and the FL can be effectively considered as a capacitor. An electric field can be applied across the ME capacitor by using terminals 1 and 2 in Fig. 1. When an electric field is applied across this ME capacitor, it exerts an effective magnetic field on the nanomagnet. If the resultant magnetic field is strong enough, it can modify the magnetization of the nanomagnet [Wu 2013]. In Fig. 1, we assume that the easy axis of the FL is in-plane (along the $x$-axis) due to shape anisotropy. When a positive voltage is applied across terminals 1 and 2, it generates an exchange bias field in the $+x$ direction. When the generated ME exchange bias field is strong enough to overcome the shape anisotropy, the FL switches in the $+x$ direction. Similarly, application of a negative voltage across terminals 1 and 2 switches the FL in the $-x$ direction. For reading the magnetization state of the nanomagnet, the FL is also used as the free layer of a magnetic tunnel junction (MTJ). As shown in Fig. 1, the MTJ consists of a magnetic pinned layer (PL), a tunneling oxide (MgO) and the magnetic free layer (FL). With the orientation of the PL shown in Fig. 1, the MTJ is in the high resistance (anti-parallel or AP) state when the FL magnetization is along $-x$ direction, and the MTJ is in low resistance state (parallel or P) when the FL magnetization is in the $+x$ direction. Therefore, the application of a strong enough positive (negative) voltage across the ME capacitor using terminals 1 and 2 can switch the MTJ to P (AP) state. The MTJ resistance can subsequently be sensed by using terminals 2 and 3 in Fig. 1. Thus, the use of ME effect not only provides an efficient mechanism for magnetization reversal, it also enables the decoupling of the read and write paths to enhance the reliability of the above ME-MTJ device operation.

The exact physical mechanism of ME effect is currently an area of intense research [Ramesh 2014, Gao 2017]. As such, we do not follow any particular experiment or material combination. Rather, we use an ME-coefficient ($\alpha_{ME}$) to abstract the efficiency of the ME effect [Nikonov 2014, Manipatruni 2015, Jaiswal 2017a, Jaiswal 2017b]. Such an abstraction is justified since our goal is not to investigate the physical origin of ME effect. We rather intend to investigate the possibility of designing an energy efficient global interconnect circuit through the use of ME-MTJ device. We next discuss the modeling of the ME-MTJ device.

### B. Modeling and Simulation Framework

Under mono-domain approximation, the magnetization dynamics of the nanomagnet can be modeled by using the Landau-Lifshitz-Gilbert equation [Gilbert 2004], which can be expressed as

$$\frac{\partial \hat{m}}{\partial t} = -|\gamma|\hat{m} \times \vec{H}_{eff} - \alpha \hat{m} \times \frac{\partial \hat{m}}{\partial t} \qquad (1)$$

where, $\hat{m}$ is the unit vector along the magnetization direction, $\alpha$ is the Gilbert damping co-efficient and $\gamma$ is the gyromagnetic ratio. $\vec{H}_{eff}$ is the effective magnetic field which can be expressed as

$$\vec{H}_{eff} = \vec{H}_{demag} + \vec{H}_{interface} + \vec{H}_{thermal} + \vec{H}_{ME} \qquad (2)$$

where, $\vec{H}_{demag}$ is the demagnetizing field, $\vec{H}_{interface}$ is the interfacial anisotropy field, $\vec{H}_{thermal}$ is the stochastic field due to thermal noise, and $\vec{H}_{ME}$ is the field due to ME effect. The demagnetization field ($\vec{H}_{demag}$) can be expressed as [Wang 2014]

$$\vec{H}_{demag} = -M_s(N_{xx}m_x\hat{x}, N_{yy}m_y\hat{y}, N_{zz}m_z\hat{z}) \qquad (3)$$

where, $m_x, m_y$ and $m_z$ are directional magnetization components, and $N_{xx}, N_{yy}$ and $N_{zz}$ are demagnetization factors which are calculated using analytical equations from [Aharoni 1998]. The interfacial anisotropy field ($\vec{H}_{interface}$) can be expressed as

$$\vec{H}_{interface} = \left(0\hat{x}, 0\hat{y}, \frac{2K_i}{\mu_0 M_s t_{FL}} m_z \hat{z}\right) \qquad (4)$$

where, $K_i$ is the interfacial anisotropy constant and $t_{FL}$ is the free layer thickness. The thermal field ($\vec{H}_{thermal}$) can be expressed through the following stochastic formalism as [Brown 1963]

$$\vec{H}_{thermal} = \vec{\zeta}\sqrt{\frac{2\alpha k_B T}{|\gamma|M_s V_{FL} dt}} \qquad (5)$$

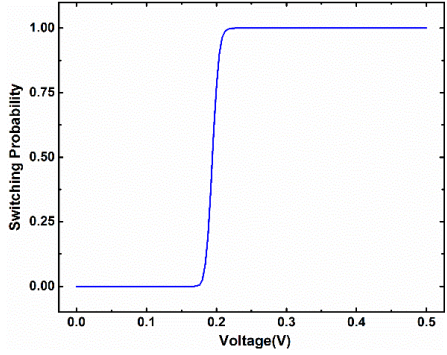

Fig. 2. Change of nanomagnet switching probability with changing voltage difference across the ME capacitor.

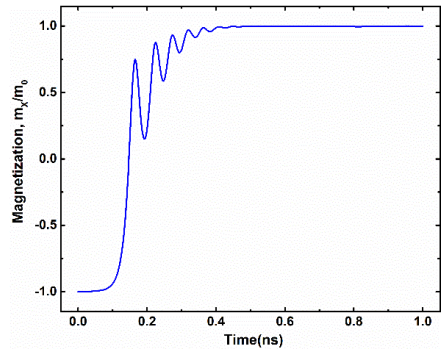

Fig. 3. Evolution of in-plane magnetization component with time of the nanomagnet for an applied voltage of $0.2\ V$ across the ME capacitor.

where, $\vec{\zeta}$ is a vector having components that are zero mean Gaussian random variables with standard deviation of 1. $V_{FL}$ is the volume of the FL and $dt$ is the simulation time-step. The ME effective field is abstracted by using the parameter $\alpha_{ME}$ as [Manipatruni 2015]

$$\vec{H}_{ME} = \left(\alpha_{ME}\left(\frac{V_{ME}}{t_{ME}}\right)\hat{x}, 0\hat{y}, 0\hat{z}\right) \quad (6)$$

where, $\alpha_{ME}$ is the ME effect coefficient, $V_{ME}$ is the applied voltage across the ME capacitor, and $t_{ME}$ is the ME oxide thickness. Equation (1) is solved numerically using Heun's method [Scholz 2001]. The solution gives the required voltage for magnetization switching along with the switching time. The device parameters used in our simulation [Jaiswal 2017b] are listed in Table 1. Using the mentioned values, we get the switching probability as a function of the voltage applied across the ME capacitor as shown in Fig. 2. When the applied voltage is greater than $\sim 0.2\ V$ for the device under consideration, it switches deterministically as shown in Fig. 2. Moreover, the complete magnetization reversal occurs within $500\ ps$. This is shown in Fig. 3, where the FL magnetization switches from $-x$ to $+x$ direction with an applied voltage of $+0.2\ V$ across the ME capacitor.

In order to model the resistance of the MTJ, we use non-equilibrium Green's function (NEGF) formalism and abstract the MTJ resistance into a behavioral model. The detail description of this method can be found in [Fong 2011]. The ME effect induced magnetization switching is incorporated with the behavioral MTJ resistance model to evaluate the MTJ resistance change. The resistance of the MTJ is then subsequently integrated with IBM $45\ nm$ CMOS technology to evaluate the interconnect circuit operations. We next present the details of our proposed global interconnect circuit.

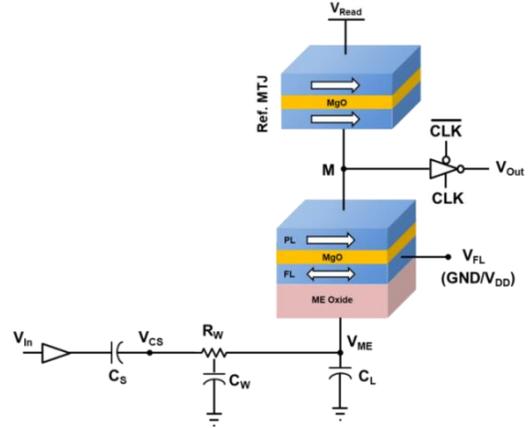

Fig. 4. Schematic of the capacitively driven global interconnect circuit with an ME-MTJ receiver.

## III. ME-MTJ AS THE RECEIVER FOR CAPACITIVELY DRIVEN LOW SWING GLOBAL INTERCONNECT

The schematic of the proposed capacitively driven global interconnect with an ME-MTJ receiver is shown in Fig. 4. A series capacitance ($C_S$) is inserted at the input side for driving the long Cu-line. $C_S$ can either be implemented as a MOS-capacitor [Mensink 2010] or a pitchfork wire capacitor [Ho 2008]. The voltage at the receiving end of the Cu-line in Fig. 4 ($V_{ME}$) can be approximated by using the capacitive divider rule as

$$V_{ME} \approx V_{IN} \frac{C_S}{C_S + C_W + C_L} \quad (7)$$

where, $C_W$ is the overall wire capacitance and $C_L$ is the capacitance at the receiving end, which is much smaller compared to the wire capacitance of a global wire. The voltage at the receiving end ($V_{ME}$) is applied to the ME oxide contact in an ME-MTJ device as shown in Fig. 4. To ensure a receiving end voltage sufficiently larger than the ME switching threshold, we use $C_S \approx C_W/2$, which gives $V_{ME} \approx 0.33\ V$ with the input amplitude of $1\ V$. When $V_{ME}$ is greater than the ME switching threshold ($\sim 0.2\ V$ for the used device dimensions), it switches the FL magnetization in the $+x$ direction (and the MTJ resistance to $R_P$) at the receiving end. The change of the MTJ resistance is sensed by using the reference MTJ as shown in Fig. 4, which creates a resistance divider network. A read current is passed through the two MTJ resistances (connected in series) by using the terminal $V_{Read}$. The read current sets the voltage at node 'M' in Fig. 4 in accordance to the resistance of the ME-MTJ. This resistive divider MTJ network drives a clocked CMOS inverter as shown in Fig. 4 to produce the appropriate digital output signal. After reading the MTJ state, the free layer is reset to $-x$ direction by applying $V_{DD}$ to the FL terminal (negative voltage across the ME capacitor) for enabling operation for the next clock cycle. Continuous operation of the above network for a $5\ mm$ long Cu line is shown in Fig. 5, which shows the applied voltage to the ME oxide ($V_{ME}$) goes above the ME switching threshold for a digital high input. The resultant change in magnetization of the free layer ($m_X/m_0$) is also shown in Fig. 5. Subsequently, the output voltage ($V_{Out}$) follows the input voltage ($V_{In}$) through the change in the ME-MTJ resistance by a digital high input as shown in Fig. 5. In addition, we performed a Monte-Carlo simulation which ensures that $V_{ME}$ always remains above the switching threshold ($0.2\,V$) for a digital high input even with 20% variation in all the device characteristics. Next, we evaluate the energy-delay performance of the proposed technique.

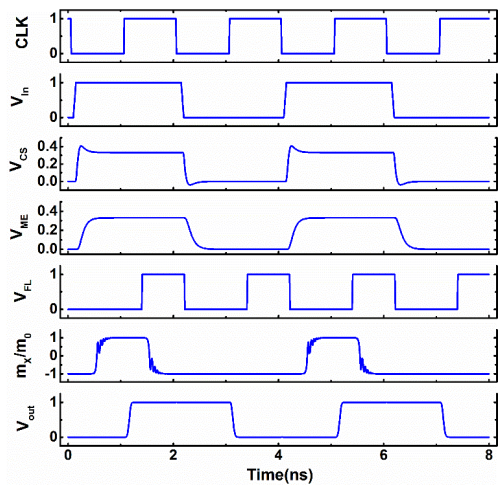

Fig. 5. Simulated waveforms for a 5 $mm$ long Cu line using the capacitive ME interconnect.

Table 2. Performance comparison with CMOS interconnects

| Cu-Line Length (mm) | Method | Energy consumption (fJ/bit/mm) | Propagation delay (ns) |
|---|---|---|---|
| 5 | Full-swing CMOS | 156.81 | 0.2673 |
|   | Low-swing capacitive CMOS | 92.11 | 0.2646 |
|   | Capacitive ME | 51.76 | 0.7673 |
| 10 | Full-swing CMOS | 155.42 | 0.5264 |
|   | Low-swing capacitive CMOS | 91.89 | 0.5153 |
|   | Capacitive ME | 47.77 | 1.0179 |

## IV. RESULTS AND DISCUSSION

Since our proposal is primarily intended for global lines, we use wider Cu-wires with lower resistance and somewhat higher capacitance for performance evaluation (unit resistance, $r_w = 50\,\Omega/mm$ and unit capacitance, $c_w = 0.25\,fF/\mu m$ [Lee 2013]). We simulate both full swing CMOS interconnect with repeaters and low swing capacitive CMOS interconnect with differential amplifier based receiver [Ho 2008] for comparison with the proposed technique. The energy consumption and propagation delay in 5 and 10 $mm$ long Cu-lines are listed in Table 2 for all the methods. The capacitive ME energy consumption shown in Table 2 includes the energy required for the read and reset operations in addition to the capacitive energy loss along the line. As shown in the table, ME receiver based capacitive low swing network offers significant improvement in energy dissipation (~$3\times$ lower than full-swing CMOS and ~$2\times$ lower than capacitive low-swing CMOS). The energy consumption in ME based design is improved because of two major factors. First, the low swing in the line (~$V_{DD}/3$) reduces the dynamic/capacitive line loss. Second, the ability to perform signal restoration without the necessity of differential signaling or amplification reduces the static/resistive dissipation. Although, low swing capacitive CMOS can reduce the signal swing in the line, the use of differential amplification for successful detection adds static energy dissipation. Full-swing CMOS design consumes the highest amount of energy because of the higher capacitive loss in the line (higher voltage swing in the line) along with additional energy loss due to the insertion of multiple repeaters. Propagation delay for the low swing capacitive CMOS design is very close to the repeated full-swing CMOS design as shown in Table 2. This is due to the use of capacitive pre-emphasis, which extends the signal bandwidth and enables signal propagation at similar speed without the need for repeaters [Ho 2008, Mensink 2010]. The delay in the wire itself is similar for capacitive ME and low swing capacitive CMOS designs. However, the overall propagation delay is higher for the capacitive ME design primarily because of the additional time required for ME effect induced magnetization switching process. Note that, the performance of capacitive ME interconnect is limited by the speed of the ME effect induced switching. ME switching process is still an active area of research and an experimental demonstration of global magnetization reversal by ME effect is yet to be available. As a result, our simulation framework is essentially a predictive model for the ME switching behavior. It has been anticipated in the literature [Manipatruni 2015] that ME effect induced magnetization reversal is likely to occur within 100 $ps$. If such limits are indeed reached, then the proposed interconnect will be able to operate much faster due to the faster ME induced switching.

It needs to be mentioned here that, spin-torque sensor based receiver for global interconnects have previously been proposed which use current induced domain-wall motion to modify an MTJ resistance [Sharad 2013, Sharad 2014, Azim 2015, Azim 2017]. Since such techniques function as current-mode interconnects, it leads to static/resistive energy dissipation along the wire. The use of capacitive ME write port in our proposal enables us to use voltage-mode interconnection, which suppresses the static dissipation. The domain-wall sensor based interconnect can operate with a lower voltage along the line compared to the proposed ME effect based interconnect. However, it requires an additional supply voltage which we can avoid through the use of a capacitive divider network.

## V. CONCLUSION

To conclude, we have exploited ME effect induced magnetization reversal to propose a capacitively driven low swing voltage-mode global interconnect. Our device-to-circuit simulation results show significant improvement in energy efficiency in addition to reduced complexity. With numerous ongoing efforts to improve ME effect induced magnetization switching process, our proposed technique can be an attractive candidate for future on-chip global signaling.

## ACKNOWLEDGMENT

This research was funded in part by C-SPIN, the center for spintronic materials, interfaces, and architecture, funded by DARPA and MARCO; the Semiconductor Research Corporation, the National Science Foundation, and the Vannevar Bush Faculty Fellows program.